\documentclass[12pt]{amsart}
\usepackage{graphicx}
\usepackage{epsfig}
\usepackage{fancyhdr}

\usepackage{amsmath,amsfonts,amssymb,amsthm,mathabx}
\theoremstyle{plain}
\newtheorem{thm}{Theorem}
\newtheorem*{thm*}{Theorem}

\newtheorem{lem}{Lemma}
\newtheorem*{lem*}{Lemma}

\newtheorem*{cor*}{Corollary}

\newtheorem{prop}{Proposition}
\newtheorem*{prop*}{Proposition}

\theoremstyle{definition}

\newtheorem{Def}{Definition}
\newtheorem{Rem}{Remark}

\newcommand{\reftit}{\textit}    
\newcommand{\refis}{\textbf}     
\def\cR{{\mathcal{R}}}
\def\cN{{\mathcal{N}}}

\def\cT{{\mathcal{T}}}

\def\be{\begin{equation}}
\def\ee{\end{equation}}
\def\ben{\begin{equation*}}
\def\een{\end{equation*}}

\begin{document}

\title{Horton Law in Self-Similar Trees}
\author{Yevgeniy Kovchegov}
\address{Department of Mathematics, Oregon State University, Corvallis, OR,  97331}
\email{kovchegy@math.oregonstate.edu}
\thanks{This research is supported by the NSF DMS-1412557 award.}

\author{Ilya Zaliapin}
\address{Department of Mathematics and Statistics, University of Nevada, Reno, NV, 89557}
\email{zal@unr.edu}



\begin{abstract}
Self-similarity of random trees is related to the operation of pruning.
Pruning $\cR$ cuts the leaves and their parental edges and removes 
the resulting chains of degree-two nodes from a finite tree.
A Horton-Strahler order of a vertex $v$ and its parental edge is defined 
as the minimal number of prunings necessary to eliminate the subtree rooted at $v$.
A branch is a group of neighboring vertices and edges of the same order.
The Horton numbers $\cN_{k}[K]$ and $\cN_{ij}[K]$ are defined as the expected number 
of branches of order $k$, and the expected number of order-$i$ branches that merged
order-$j$ branches, $j>i$, respectively, in a finite tree of order $K$.
The Tokunaga coefficients are defined as $T_{ij}[K]=\cN_{ij}[K]/\cN_{j}[K]$.
The pruning decreases the orders of tree vertices by unity.
A rooted full binary tree is said to be mean-self-similar if its Tokunaga
coefficients are invariant with respect to pruning: 
$T_k:=T_{i,i+k}[K]$. 
We show that for self-similar trees, the condition
$\limsup_{k\to\infty} T_k^{1/k}<\infty$ is necessary
and sufficient for the existence of the strong Horton law:
$\cN_{k}[K]/\cN_{1}[K]\to R^{1-k}$, as $K\to\infty$ for some $R>0$ and every $k\ge 1$.
This work is a step toward providing rigorous foundations for the Horton law that, 
being omnipresent in natural branching systems, has escaped so far a formal explanation. 
\end{abstract}



\date{\today}
\maketitle

\section{Introduction}
\label{intro}
Horton laws, which are akin to a power-law distribution of the element sizes in 
a branching system, epitomize the scale invariance of natural dendritic structures.
It is very intuitive that the existence of Horton laws should be related to the
self-similar organization of branching, defined in suitable terms.
Such relation, however, has escaped a rigorous explanation, remaining for long time
a part of science literature folklore (e.g., \cite{Kirchner93,DR99}). 
This paper shows that a weak (mean) invariance under the operation of tree pruning is sufficient
for the Horton law of branch numbers to hold in the strongest sense, hence 
explaining and unifying many earlier empirical observations and partial results 
in this direction.

We work with binary trees, although our results can be 
easily extended to the case of trees of a higher degree.
Recall that {\it pruning} of a finite rooted full binary tree $T$ cuts its leaves 
(vertices of degree one) and their parental edges, and removes the resulting
chains of degree-two vertices and their parental edges (so-called series reduction).
The {\it Horton-Strahler order} $k$ of vertex $v$ is the minimal number of prunings
necessary to eliminate the subtree rooted at $v$.
A {\it branch} is a sequence of neighboring vertices and their 
parental edges of the same order.
We write $N_k$ for the total number of branches of order $k$ in a tree.
A common empirical observation in the natural dendritic structures is 
that $N_{k+1}/N_k\approx R$, $3\le R \le 5$. 
This regularity was first described by Robert E. Horton \cite{Horton32,Horton45} 
in a study of river streams; it has been strongly corroborated in hydrology
\cite{Shreve66,Kirchner93,Pec95,Tarboton96,GW98,RR97} and expanded 
to biology and other areas \cite{NTG97} since then.
Similar relations, referred to as {\it Horton laws}, are reported for selected 
metric quantities, 
for example the average lengths of river streams ($l_{k+1}/l_k\approx R_l$), average 
contributing areas of order-$k$ drainage basin ($A_{k+1}/A_k\approx R_A$), etc.

Informally, Horton laws suggest that the branch order $k$ is proportional to the 
logarithm of a suitably defined ``size'' $S_k$ of the branch: $S_k \propto B^k$. 
A geometric distribution of the branch counts $N_k$,
\[{\sf P}\left({\rm a~random~branch~has~order}\ge k\right)=R^{-k},\]
is equivalent to a power-law distribution of branch sizes: 
\[{\sf P}\left(S_k \ge x\right) \propto\,x^{-\alpha},\quad\alpha=\ln(R)/\ln(B).\]
Hence, the empirical Horton laws can be interpreted as a power-law distribution
of system element sizes. 
This might hint at a scale-invariant organization of the respective branching structures,
as power laws often accompany fractality.

For long time, the only rigorous result on validity of Horton laws was
that of Ronald Shreve \cite{Shreve67}, who demonstrated that in a uniform distribution 
of rooted binary trees with $n$ leaves (that he called {\it topologically
random networks}), the ratio $N_{k+1}/N_k$ converges to 4 as $n$ goes to 
infinity.
This model is equivalent to the critical binary Galton-Watson tree 
conditioned to have $n$ leaves (e.g., \cite{Pitman, BWW00}).
Shreve \cite{Shreve69} also showed that in a topologically
random network the average number $T_{ij}$ of side-branches of order $i$
per branch of order $j$ only depends on the relative ordering
of the branches: $T_{ij}=2^{j-i-1}$, as the tree size increases.
We notice that pruning decreases the order of every branch by unity.
Accordingly, Shreve's result implies, in particular, that the average numbers $T_{ij}$
are invariant under the pruning operation: $T_{ij}=T_{i-1,j-1}$.
The topologically random network was hence the first example of a model 
that obeys both the Horton law of branch numbers and structural invariance 
with respect to pruning.
The invariance with respect to pruning is called {\it self-similarity},
and may refer to the invariance of distributions, or the means of selected
statistics (like is the case with Shreve's result).

Eiji Tokunaga \cite{Tok78} introduced a broader class of mean-invariant models
defined by the constraint $T_{ij} = T_{j-i} = a\,c^{j-i-1}$ for positive $a,c$.
The validity of the Tokunaga constraint has been empirically confirmed in numerous
observed and modeled systems (see \cite{MT93,Tarboton96,NTG97,TPN98,ZZF13} and references therein),
notably including diffusion limited aggregation \cite{Oss92,MT93}, and 
two dimensional site percolation \cite{TMM+99,YNT+05,ZWG06a}.
Furthermore, Burd, Waymire, and Winn \cite{BWW00} demonstrated that the Tokunaga
constraint with $(a,c)=(1,2)$ is the characteristics property of critical
binary offspring distribution within the class of Galton-Watson (non necessarily binary) 
trees, and that the critical binary Galton-Watson trees are also distributionally
invariant with respect to pruning. 
Zaliapin and Kovchegov \cite{ZK12} have shown that both Horton law 
with $R=4$ and the Tokunaga constraint with $(a,c)=(1,2)$ hold in a 
level-set tree representation of a symmetric random walk, and that in general
such a tree is not equivalent to the critical binary Galton-Watson model.

McConnell and Gupta \cite{MG08} have shown that the Tokunaga constraint is sufficient 
for a Horton law.
Specifically, they proved that if the sequence of branch counts $N_k$ is related
to the Tokunaga coefficients $T_k=a\,c^{k-1}$ via the recursive counting equation
\be
\label{count}
N_k = 2\,N_{k+1}+\sum_{j=1}^{K-j}T_j\,N_{k+j},\quad 1\le k\le K-1, \quad K\ge 2,
\ee
then $N_{k+1}/N_k \to R$ for any $k\ge 1$ in
the limit of large-order trees. 
In this case
\be
\label{R1}
R = \frac{2+a+c+\sqrt{(2+a+c)^2-8\,c}}{2},
\ee
which was reported earlier (under the explicit assumption that Horton law holds)
by Tokunaga \cite{Tok78}, Peckham \cite{Pec95}, and others.
 
The equation \eqref{R1} suggests that different Horton
exponents $R$ can be easily attained by using the Tokunaga side-branching with
different pairs $(a,c)$ (see, e.g. \cite{NTG97}).
At the same time, most of the existing rigorous results on the Horton laws in
``natural'' models (not formulated explicitly in terms of Horton branch counting) refer 
to the models equivalent to the Galton-Watson critical binary tree or its slight 
ramifications, with $R=4$, or to trees with no side-branching and $R=2$.
Recently, the authors established a weak version of the Horton law for the tree
that describes the celebrated Kingman's coalescent process; this system has $R = 3.043827\dots$
\cite{KZ15}.

This study expands the sufficient conditions for the Horton law 
(in its strong version defined in Sect.~\ref{hl}) 
to all sequences of side-branch coefficients such that $\limsup_k T_k^{1/k}<\infty$.
We also show that this condition is necessary in the class of mean self-similar
trees.
The Horton exponent in this case is given by $R=1/w_0$, where $w_0$ is the
only real root of 
\[\hat t(z) = -1 + 2z +\sum_{k=1}^{\infty} z^k\,T_k\]
within the interval $[0,1/2]$, which was conjectured by Peckham \cite{Pec95}.
The results are obtained in a probabilistic setting and refer to the
expectations of branch counts with respect to a probability
measure on the space of finite rooted full binary trees of Horton-Strahler 
order $K$, as $K$ increases. 
This set-up allows us to relax the assumption of similar statistical structure
of side-branching within each branch, which is a typical assumption
in the studies of Horton laws and tree self-similarity \cite{MG08,Pec95}.

We start by reviewing the essential definitions in Sect.~\ref{prelim}.
Section \ref{sst} introduces self-similar trees in a probabilistic setting,
and establishes the equivalence of prune-invariance and the 
constraint $T_{ij}=T_{j-i}$.
The main results are proven in Sect.~\ref{results}.

\section{Preliminaries}
\label{prelim}

\subsection{Rooted trees}
\label{rt}
Recall that a simple graph is a collection of vertices
connected by edges in such a way that each pair of vertices
may have at most one connecting edge and there is no 
self-loops.
A {\it tree} is a connected simple graph without cycles.
In a {\it rooted} tree, one node is designated as a root; this
imposes the parent-child relationship between the neighbor vertices. 
Specifically, of the two neighbor vertices the one closest to
the root is called {\it parent}, and the other -- {\it child}.
In a rooted tree each non-root vertex has the unique {\it parental} 
edge that connects this vertex to its parent.
A {\it leaf} is a vertex with no children.
The space of finite unlabeled rooted full binary trees, including
the empty tree $\phi$, is denoted by $\cT$.
All internal vertices in a tree from $\cT$ have degree 3, leaves
have degree 1, and the root has degree 2. 

\subsection{Tree pruning}
\label{pruning}
{\it Pruning} of a tree is an onto function $\cR:\cT\to\cT$,
whose value $\cR(T)$ for a tree $T\ne\phi$ is obtained by removing
the leaves and their parental edges from $T$, and then compressing the 
resulting tree from $\tilde\cT$ by removing all degree-two chains (this
operation is known as {\it series reduction}).
We also set $\cR(\phi)=\phi$.

\subsection{Horton-Strahler orders}
\label{hst}
\begin{figure}[t] 
\centering\includegraphics[width=\textwidth]{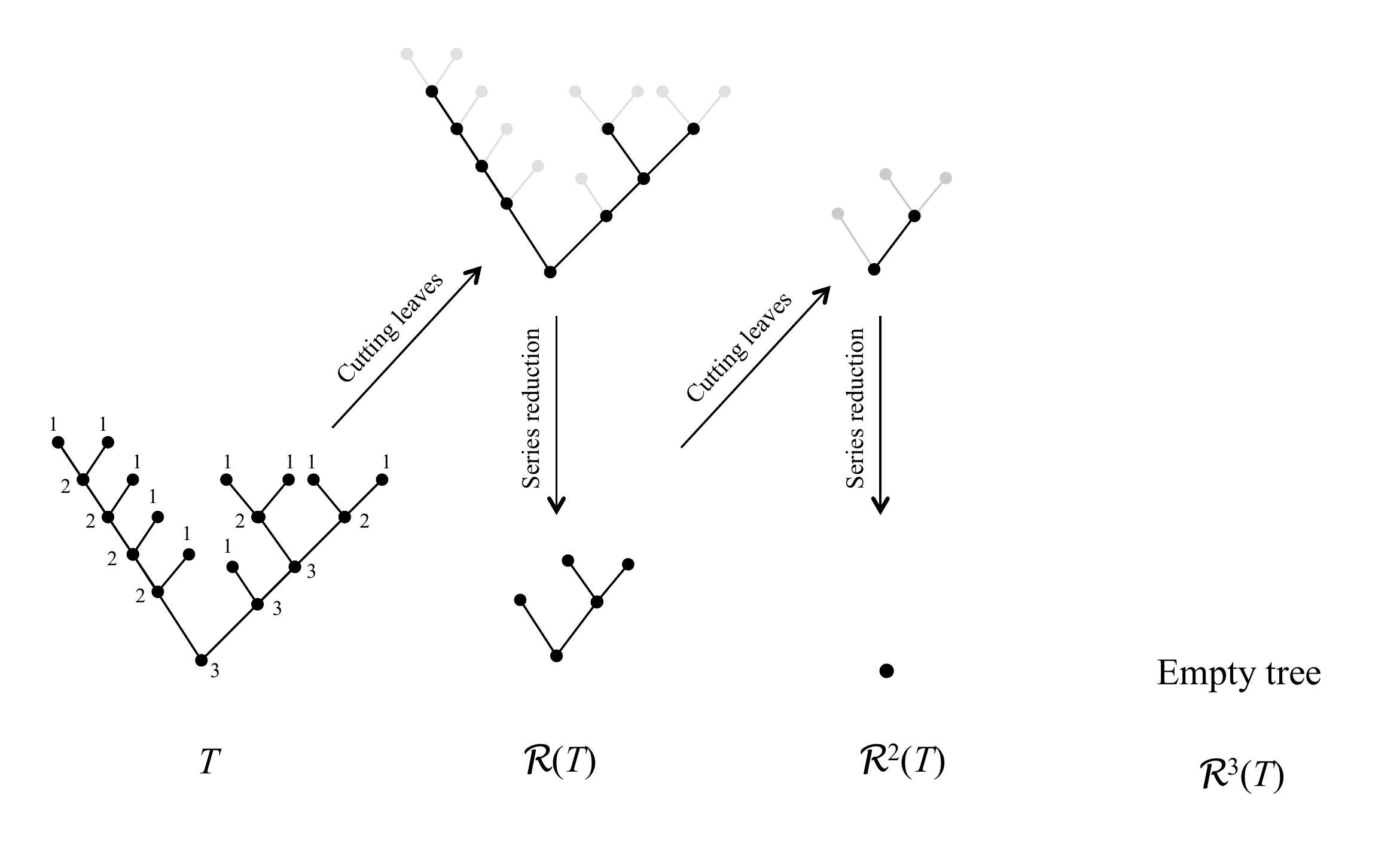}
\caption[Example of Horton-Strahler indexing]
{Example of pruning and Horton-Strahler ordering.
The Horton-Strahler orders are shown next to each vertex
of the initial tree $T$. The figure shows the two stages of each pruning --
cutting the leaves (top row), and consecutive series reduction (bottom row).
The order of the tree is $k(T)=3$ with $N_1=10$, $N_2=3$, $N_3=1$, and $N_{1,2}=3$,
$N_{1,3}=1$, $N_{2,3}=1$. }
\label{fig_HST}
\end{figure}

The Horton-Strahler ordering of the vertices and edges of a finite
rooted binary tree $T\in\cT$ is related to the iterations $\cR^k$ of the pruning
operation \cite{Horton45,Strahler,Pec95}. 
Specifically, a vertex $v\in T$ and its parental edge have order $k=1,2,\dots$ if the
subtree $\tau_v\in T$ rooted at $v$ is eliminated during the $k$-th iteration of 
pruning:
\[k(v)=\min_{k\ge 1}\left(\cR^k(\tau_v)=\phi\right).\]
The order $k(T)$ of a non-empty tree coincides with the maximal order of its vertices.
We also set $k(\phi)=0$.
A {\it branch} is defined as a union of neighboring vertices and edges of
the same order.
Figure~\ref{fig_HST} illustrates the operation of pruning and the definition
of Horton-Strahler orders.

Equivalently, the Horton-Strahler ordering can be done by hierarchical 
counting \cite{Pec95,NTG97,BWW00}.
In this approach, each leaf is assigned order $k({\rm leaf})=1$.
An internal vertex $p$ whose children have orders $i$ and $j$
is assigned the order 
\[k(p)=\max\left(i,j\right)+\delta_{ij},\] 
where $\delta_{ij}$ is the Kronecker's delta.
The parental edge of a vertex has the same order as the vertex.

\subsection{Horton law}
\label{hl}
Let $\cT_{K}$, $K\ge 1$, be the subspace of finite binary trees of Horton-Strahler order $K$.
Consider a set of probability measures $\{\mu_{K}\}_{K\ge 1}$, each of which is defined
on $\cT_{K}$, and write ${\sf E}_K(\cdot)$ for the mathematical expectation with respect to $\mu_{K}$.
Let $N_k=N_k[T]$ be the number of branches of order $k$ in a tree 
$T\in\cT$.
We define the average Horton numbers, which are the main object for our analysis: 
\[\cN_{k}[K]= {\sf E}_K(N_k),\quad 1\le k\le K,\quad K\ge 1.\]

\begin{Def}
\label{Horton}
We say that a sequence of measures $\{\mu_K\}$ satisfies a {\it strong Horton law} if
\[\lim_{K\to\infty} \frac{\cN_{k}[K]}{\cN_{1}[K]}= R^{1-k} < \infty\quad {\rm for~any~}k\ge1.\]
\end{Def}

\subsection{Tokunaga coefficients}
Let $N_{ij}=N_{ij}[T]$ denote the number of instances when an order-$i$ 
branch merges with an order-$j$ branch, $1\le i <j$, in a tree $T$.
Such branches are referred to as {\it side-branches} of order ${ij}$.
Define the respective expectation $\cN_{ij}[K]={\sf E}_K(N_{ij})$.
The {\it Tokunaga coefficients} $T_{ij}[K]$ for subspace $\cT_{K}$ are defined as
\be
\label{def_tok}
T_{ij}[K]=\frac{\cN_{ij}[K]}{\cN_{j}[K]}, \quad 1\le i <j\le K.
\ee

\begin{Rem}
\label{rem}
Consider a situation when every branch of order $j$ has the same 
expected number $S_{ij}$ of side-branches of order $i<j$.
Then
\[\cN_{ij}[K]={\sf E}_K(N_{ij})={\sf E}_K\left({\sf E}_K(N_{ij}|N_j)\right)
={\sf E}_K(N_j\,S_{ij})=S_{ij}\,{\sf E}_K(N_j)=S_{ij}\,\cN_j[K],\]
and hence
\[T_{ij}[K]=\frac{\cN_{ij}[K]}{\cN_{j}[K]}=S_{ij}.\]
Such framework was considered by Shreve \cite{Shreve69}, Tokunaga \cite{Tok78}, 
Burd, Waymire and Winn \cite{BWW00} and others.
Our definition \eqref{def_tok} includes this situation as a special case, although in general it is 
free of the assumption of similar statistical structure of individual branches. 
\end{Rem}

\section{Self-similar trees}
\label{sst}

\begin{Def}
\label{coord}
A set of measures $\{\mu_K\}$ on $\{\cT_K\}$ is called {\it coordinated}
if $T_{ij}:=T_{ij}[K]$ for all $K\ge 2$ and $1\le i< j\le K$.
\end{Def}
\noindent For a set of coordinated measures $\{\mu_K\}$, the Tokunaga matrix $\mathbb{T}_K$ for any $K$ 
forms a $K \times K$ matrix
$$\mathbb{T}_K=\left[\begin{array}{ccccc}0 & T_{1,2} & T_{1,3} & \hdots & T_{1,K} \\0 & 0 & T_{2,3} & \hdots & T_{2,K} \\0 & 0 & \ddots & \ddots & \vdots \\\vdots & \vdots & \ddots & 0 & T_{K-1,K} \\0 & 0 & 0 & 0 & 0\end{array}\right],$$
which coincides with the restriction of any larger-order Tokunaga matrix $\mathbb{T}_M$, $M>K$,
to the first $K\times K$ entries.

\begin{Def}
\label{ss1}
A collection of coordinated  probability measures $\{\mu_{K}\}$ on $\{\cT_K\}$ is called 
(mean) {\it self-similar} if $T_{ij} = T_{j-i}$ for some sequence $T_k\ge 0$, $k=1,2,\dots,$ 
and any $K\ge 2$.
The elements of the sequence $T_k$ are also referred to as Tokunaga coefficients, 
which does not create confusion with $T_{ij}$.
\end{Def}
\noindent For a self-similar collection of measures the matrix of Tokunaga coefficients becomes Toeplitz:
$$\mathbb{T}_K=\left[\begin{array}{ccccc}0 & T_1 & T_2 & \hdots & T_{K-1} \\0 & 0 & T_1 & \hdots & T_{K-2} \\0 & 0 & \ddots & \ddots & \vdots \\\vdots & \vdots & \ddots & 0 & T_1 \\0 & 0 & 0 & 0 & 0\end{array}\right].$$

A variety of self-similar measures can be constructed for 
an arbitrary sequence of Tokunaga coefficients $T_k>0$, $k\ge 1$.
Next, we give one natural example.

\medskip

{\bf Example 1: Independent Random Attachment.}
The subspace $\cT_1$, which consists of a single-vertex tree,
possess a trivial unity mass measure.
To construct a random tree from $\cT_2$, we select a discrete probability
distribution $P_{1,2}(n)$, $n=0,1,\dots$, with the mean value $T_1$.
A random tree $T\in\cT_2$ is obtained from the single-vertex 
tree $\tau_1$ of order 1 via the following two operations.
First, we attach two child vertices to the only vertex of $\tau_1$.
This creates a tree of order 2 with no side-branches -- two leaves attached to the root.
Second, we draw the number $N_{1,2}$ from the distribution $P_{1,2}$, and
attach $N_{1,2}$ vertices to this tree so that they form side-branches 
of order $\{1,2\}$.

In general, to construct a random tree from $\cT_{K}$ for $K\ge 2$ we select a set 
of discrete probability distributions $P_{k,K}(n)$, $k=1,...,K-1$, with 
the respective mean values $T_k$.
A random tree $T\in\cT_{K}$ is constructed in iterative fashion, starting from
the single-vertex tree $\tau_1$ and increasing its order by adding new vertices.
Specifically, to construct a random tree $\tau_k$ of order $k\ge 2$ from a random tree $\tau_{k-1}$ 
of order $k-1$, we perform the following operations.
First, add two new child vertices to every leaf of $\tau_{k-1}$ hence producing
a tree $\tilde\tau_k$ of order $k$ with no side-branches of order 1.
Second, for each branch of order $2\le j\le k$ in $\tilde\tau_k$ draw a random number $N_{1j}$ 
from the distribution $P_{j-1,K}$ and attach $N_{1j}$ new child vertices to 
this branch so that they form side-branches of order 1. 
Each new vertex is attached in random order with respect to
the existing side-branches.
Specifically, we notice that $s\ge 0$ side-branches attached to a branch of order $j$
are uniquely associated with $s+1$ edges within this branch. 
(When discussing the single branch of the maximal order $k$, we 
count one ``imaginary'' edge parental to the tree root.)
The attachment of the new $N_{1j}$ vertices among the $s+1$ edges
is given by the equiprobable multinomial distribution with $s+1$ categories
and $N_{1j}$ trials.

According to Remark \ref{rem}, the self-similarity condition $T_{i,i+k}[K] = T_{k}$ holds 
within each subspace $\cT_{K}$, $K\ge 2$.

\medskip

Notice that pruning defines a down-shift of the order subspaces, that is for $K\ge 1$
\[\cR(\cT_K) = \cT_{K-1}.\]
Moreover, pruning
decreases the Horton-Strahler order of each vertex (and hence of each branch) by unity; 
in particular
\be
\label{shift1}
N_k[T] = N_{k-1}\left[\cR(T)\right],\quad k\ge 2,
\ee
\be
\label{shift2}
N_{ij}[T] = N_{i-1,j-1}\left[\cR(T)\right],\quad 2\le i<j.
\ee
This shift property allows us to establish connection between the values of Tokunaga 
coefficients for different orders $K$.
Specifically, consider measure $\mu^{\cR}_K$ induced on $\cT_K$ by the pruning operator:
\[\mu^{\cR}_K(A) = \mu_{K+1}\left(\cR^{-1}(A)\right)\quad \forall A\subset \cT_K.\]
The Tokunaga coefficients computed on $\cT_K$ using the induced measure
$\mu^{\cR}_K$ are denoted by $T_{ij}^{\cR}[K]$.  

\begin{Def}
\label{ss2}
A collection of coordinated probability measures $\{\mu_K\}$ on $\{\cT_K\}$ is called 
{\it self-similar} if $T_{ij}[K] = T_{ij}^{\cR}[K]$
for any $K\ge 2$ and all $1\le i < j\le K$.
\end{Def}

\begin{lem}
\label{equiv12}
The Definitions \ref{ss1},\ref{ss2} are equivalent.
\end{lem}
\begin{proof}
The pruning-related index shift \eqref{shift1},\eqref{shift2}
implies
\be
\label{shift}
T_{i+1,j+1}[K+1] = T_{ij}^{\cR}[K].
\ee

\noindent
$\bullet [\ref{ss1}\Rightarrow\ref{ss2}]:$ If a coordinated set of measures $\{\mu_K\}$ satisfy Definition \ref{ss1}, then
\[T_{i+1,j+1}[K+1] \stackrel{{\rm Def}~\ref{ss1}}{=}
T_{j-i} \stackrel{{\rm Def}~\ref{ss1}}{=}
T_{ij}[K+1] \stackrel{{\rm coordination}}{=} T_{ij}[K].\]
Together with \eqref{shift}, this implies 
\[T_{ij}[K] = T_{ij}^{\cR}[K],\]
which means that Definition \ref{ss2} is also satisfied.

\noindent
$\bullet [\ref{ss2}\Rightarrow\ref{ss1}]:$ If a coordinated set of measures $\{\mu_K\}$ satisfy Definition \ref{ss2}, then 
\[T_{ij}[K] \stackrel{{\rm Def~\ref{ss2}}}{=} 
T_{ij}^{\cR}[K]\stackrel{{\rm by~} \eqref{shift}}{=} 
T_{i+1,j+1}[K+1]\stackrel{{\rm coordination}}{=}T_{i+1,j+1}[K].\]
Hence
\[T_{i+1,j+1}[K] = T_{ij}[K] = \dots = T_{1,j-i+1}[K]=:T_{j-i},\]
which means that Definition \ref{ss1} is also satisfied.

\end{proof}

\section{Results}
\label{results}
Consider a set of self-similar measures $\{\mu_K\}_{K\ge 1}$ with Tokunaga 
coefficients $\mathbb{T}_K$. 
We define the vector of Horton indices as
\[\zeta_K=\left(\begin{array}{c}\cN_{1}[K] \\\cN_{2}[K] \\\vdots \\ \cN_{K}[K]\end{array}\right). \]
We also define the vector of normalized Horton indices in $\mathbb{R}^\infty$,
$$\xi_K:={1 \over \zeta_K(1)}\left(\begin{array}{c}\zeta_K \\0 \\0 \\\vdots \end{array}\right)=\left(\begin{array}{c}1 \\\cN_{2}[K]/\cN_{1}[K] \\\vdots  \\\cN_{K}[K]/\cN_{1}[K]\\0 \\0 \\\vdots \end{array}\right).$$

The average number of side-branches of order $1\le i<K$ within $\cT_K$ is
$\cN_{i}[K]-2\cN_{i+1}[K]$.
At the same time, the number of side-branches of order $i$ can be computed
by counting the side-branches of order $i$ for all larger-order branches:
$$\sum\limits_{j=i+1}^K T_{ij}\,\cN_{j}[K]=\sum\limits_{m=1}^{K-i}T_m\,\cN_{i+m}[K],$$
and therefore the vector of side-branches is $\mathbb{T}_K \zeta_K$.  
Thus
\begin{equation}\label{TKev}
\mathbb{T}_K \zeta_K=\left[\begin{array}{ccccc}1 & -2 & 0 & \hdots & 0 \\0 & 1 & -2 & \ddots & \vdots \\0 & 0 & \ddots & \ddots & 0 \\\vdots & \vdots & \ddots & 1 & -2 \\0 & 0 & \hdots & 0 & 0\end{array}\right]\zeta_K.
\end{equation}
This also can be written as
\be
\label{count1}
\cN_{k}[K] = 2\,\cN_{k+1}[K]+\sum_{j=1}^{K-j}T_j\,\cN_{k+j}[K],\quad 1\le k\le K-1, \quad K\ge2,
\ee
which is a probabilistic (mean) version of a deterministic counting equation \eqref{count}.

Next, define
$$\mathbb{G}_K:=\left[\begin{array}{ccccc}-1 & T_1+2 & T_2 & \hdots & T_{K-1} \\0 & -1 & T_1+2 & \hdots & T_{K-2} \\0 & 0 & \ddots & \ddots & \vdots \\\vdots & \vdots & \ddots & -1 & T_1+2 \\0 & 0 & 0 & 0 & -1\end{array}\right]$$
which is a $K \times K$ restriction of the following infinite dimensional linear operator to the first $K$ dimensions:
\begin{equation} \label{gen}
\mathbb{G}:=\left[\begin{array}{ccccc}-1 & T_1+2 & T_2 & T_3 & \hdots  \\0 & -1 & T_1+2 & T_2 & \hdots  \\0 & 0 & -1 & T_1+2 & \ddots  \\0 & 0 & 0 & -1  & \ddots \\\vdots & \vdots & \ddots & \ddots  & \ddots \end{array}\right].
\end{equation}
Equation (\ref{TKev}) implies $\mathbb{G}_K \zeta_K = -e_K$, the $K$-th coordinate basis vector, and therefore
\begin{equation}\label{Ninv}
\left(\begin{array}{c}\zeta_{K+1}(2) \\\zeta_{K+1}(3)\\\vdots \\\zeta_{K+1}(K+1) \end{array}\right)=-\mathbb{G}_K^{-1}e_K =\left(\begin{array}{c}\zeta_K(1) \\\zeta_K(2) \\\vdots \\\zeta_K(K) \end{array}\right).
\end{equation}
Thus we proved the following.
\begin{prop}\label{prop1}
Let $\{\mu_K\}$ be a set of self-similar measures on $\{\cT_K\}$.
Then for any $K\ge 1$ and $1 \leq j \leq K$,
\[\cN_{j+1}[K+1]=\zeta_{K+1}(j+1)=\zeta_K(j)=\cN_{j}[K].\]
Accordingly, we also have
\[\cN_{i+1,j+1}[K+1] = \cN_{ij}[K], \quad 1\le i<j\le K,\quad K\ge 2.\]
\end{prop}

\bigskip
\noindent
Observe that $\mathbb{G} \xi_K = {-1 \over \zeta_K(1)} e_K$, where, by construction, 
$~\zeta_K(1)=\cN_{1}[K] \geq (T_1+2)^{K-1}$. 
The following proposition formalizes the condition required for $\lim\limits_{K \rightarrow \infty} \xi_K = \xi$, where $\xi$ satisfies $\mathbb{G} \xi=0$ with coordinates $\xi(j)=R^{1-j}$. Finally, Theorem \ref{thm_cor} at the end of this section provides a complete analysis of  $\lim\limits_{K \rightarrow \infty} \xi_K$ in terms of the sequence $T_j$ of Tokunaga coefficients.

\begin{prop} \label{main}
Let $\{\mu_K\}$ be a set of self-similar measures on $\{\cT_K\}$.
Suppose that the limit
\be
\label{R}
R=\lim\limits_{K\rightarrow\infty} {\zeta_{K+1}(1)\over \zeta_K(1)}
=\lim\limits_{K \rightarrow \infty} {\cN_{1}[K+1] \over \cN_{1}[K]}
\ee
exists and is finite.
Then, the strong Horton law holds; that is, for each positive integer $j$
$$\xi(j)=\lim\limits_{K \rightarrow \infty}\xi_K(j) = 
\lim\limits_{K\to\infty} \frac{\cN_{j}[K]}{\cN_{1}[K]}=R^{1-j}.$$
Conversely, if the limit \eqref{R} does not exist, then neither will
$\lim\limits_{K \rightarrow \infty}\xi_K(j)$. That is, the limit $\lim\limits_{K \rightarrow \infty}\xi_K(j)$ 
does not exist at least for some $j$.
\end{prop}

\begin{proof}
Suppose that the limit $R=\lim\limits_{K\rightarrow\infty} {\zeta_{K+1}(1)\over \zeta_K(1)}$ exists and is finite.
Proposition \ref{prop1} implies for any fixed integer $m\geq 1$,
$${\zeta_K(m+1) \over \zeta_K(m)}={\zeta_{K-m}(1) \over \zeta_{K-m+1}(1)} \rightarrow R^{-1}.$$
Thus, for any fixed integer $j\geq 2$,
$$\xi_K(j)={\zeta_K(j) \over \zeta_K(1)}=\prod\limits_{m=1}^{j-1} {\zeta_K(m+1) \over \zeta_K(m)} \rightarrow  R^{1-j}.$$

Conversely, suppose the limit $\lim\limits_{K\rightarrow\infty} {\zeta_{K+1}(1)\over \zeta_K(1)}$ does not exist. Then, taking $j=2$, we obtain
$$\xi_K(2)={\zeta_K(2) \over \zeta_K(1)}={\zeta_{K-1}(1) \over \zeta_K(1)}$$
by Proposition \ref{prop1}. Thus $\lim\limits_{K \rightarrow \infty}\xi_K(2)$ diverges.

\end{proof}

\begin{Rem}
\label{rem1}
The conditions of Proposition~\ref{main} can be somewhat relaxed.
Specifically, we only use the self-similarity requirement to prove the 
shift equality \eqref{Ninv}.
Hence, this equality together with the existence of the limit \eqref{R}
suffice to obtain the strong Horton law.
\end{Rem}

\subsection{Expressing $\zeta_K(1)$ from $\{T_j\}$} 
In this section we express $\zeta_K(1)$ in terms of 
the elements of the Tokunaga sequence $\{T_j\}_{j=1,2,\hdots}$,
under the assumption of a ``tamed'' Tokunaga sequence:
$~\limsup\limits_{j \rightarrow \infty} T_j^{1/j} <\infty$. 
We define $$t(i)=\begin{cases}
      -1 & i=0 \\
      T_1+2 & i=1 \\
      T_i & i \geq 2
\end{cases}$$
and let $\hat{t}(z)=\sum\limits_{j=0}^\infty z^j t(j)=-1+2z+\sum\limits_{j=1}^\infty z^j T_j$. 
The quantity $\zeta_K(1)$ can be computed by counting, and expressed via convolution products as follows:
\begin{eqnarray*}
\zeta_{K+1}(1) & = & \sum\limits_{r=1}^K \sum\limits_{\substack{j_1,j_2,\hdots,j_r \geq 1\\ j_1+j_2+\hdots+j_r =K}}t(j_1)t(j_2)\hdots t(j_r)\\
& =  & \sum\limits_{r=1}^K \underbrace{(t+\delta_0)\ast (t+\delta_0) \ast \hdots \ast (t+\delta_0)}_\text{r \text{ times }}(K)\\
& =  & \sum\limits_{r=1}^\infty \underbrace{(t+\delta_0)\ast (t+\delta_0) \ast \hdots \ast (t+\delta_0)}_\text{r \text{ times }}(K),\\
\end{eqnarray*}
where $\delta_0(j)$ is the Kronecker delta, and therefore, $(t+\delta_0)(0)=0$. Hence, taking the \mbox{$z$-transform} of $\zeta_{K+1}(1)$, we obtain
\begin{equation} \label{Tzeta}
\sum_{K=1}^\infty z^{K-1}\zeta_K(1)=1+\sum\limits_{r=1}^\infty \Big[\widehat{(t+\delta_0)}(z)\Big]^r=1+\sum\limits_{r=1}^\infty \Big[\hat{t}(z)+1\Big]^r=-{1 \over \hat{t}(z)}
\end{equation}
for $|z|$ small enough.

\bigskip
\noindent
For a holomorphic function expanding in a power series $f(z)=\sum\limits_{j=0}^\infty a_j z^j$ in a nonempty neighborhood of zero containing  $|z|\leq \rho$, define $\check{f}(j)={1 \over 2\pi i}\oint\limits_{|z|=\rho} {f(z) \over z^{j+1}} dz=a_j$. Then we arrive with the following formula, expressing $\zeta_K(1)$ from $\{T_j\}$,
\begin{equation} \label{zetaT}
\zeta_{K+1}(1)=
-\widecheck{\left(\frac{1}{\widehat ~ \!\! t}\right)}(K). 
\end{equation}

\bigskip
\noindent
\begin{lem}\label{minmod}
Let $w_0$ be the only real root of  $\hat{t}(z)=-1+2z+\sum\limits_{j=1}^\infty z^j T_j$ in the interval $\left(0,{1 \over 2}\right]$. Then, for any other root $w$ of  $~\hat{t}(z)$, we have $|w|>w_0.$
\end{lem}
\begin{proof}
Observe that  since $\{T_j\}$ are all nonnegative reals, $\overline{\hat{t}(\bar{z})}=\hat{t}(z)$, and that the radius of convergence of $\sum\limits_{j=1}^\infty z^j T_j$ must be greater than $w_0$. Suppose $w=re^{i\theta}$ \mbox{($0 \leq \theta <2\pi$)} is a root of magnitude at most $w_0$. That is $~\hat{t}(w)=0~$ and $$r:=|w| \leq w_0.$$ Then $~\hat{t}(\bar{w})=0~$ and
$$0 = {1 \over 2}\Big[\hat{t}(w)+\hat{t}(\bar{w})\Big]=-1+2r\cos(\theta)+\sum\limits_{j=1}^\infty r^j T_j \cos(j\theta)$$
If $r<w_0$, then
$$0 = -1+2r\cos(\theta)+\sum\limits_{j=1}^\infty r^j T_j \cos(j\theta)\leq -1+2r+\sum\limits_{j=1}^\infty r^j T_j  < -1+2w_0+\sum\limits_{j=1}^\infty w_0^j T_j =0$$
arriving to a contradiction. Thus $r=w_0$.

\medskip
\noindent
Next we show that $\theta=0$. Suppose not. Then
$$0 = -1+2r\cos(\theta)+\sum\limits_{j=1}^\infty r^j T_j \cos(j\theta) < -1+2r+\sum\limits_{j=1}^\infty r^j T_j  = -1+2w_0+\sum\limits_{j=1}^\infty w_0^j T_j =0$$
arriving to another contradiction.
Hence $r=w_0$, $\theta=0$, and $w=w_0$.
\end{proof}

\bigskip
\noindent
Let $w_0$ denote the only root of  $\hat{t}(z)$ in the real line subinterval $\left(0,{1 \over 2}\right]$ as in Lemma \ref{minmod}.  
Recall that the radius of convergence $L=\left( \limsup\limits_{j \rightarrow \infty} T_j^{1/j}\right)^{-1}$ of $~\hat{t}(z)=-1+2z+\sum\limits_{j=1}^\infty z^j T_j$ is greater than $w_0$. Then, following Lemma \ref{minmod}, there is a positive real $\gamma \in (w_0,L)$ such that  
\begin{equation}\label{eq:gamma}
\gamma<w ~\text{ for all }~w \not= w_0~\text{ such that }~\hat{t}(w)=0.
\end{equation}
Now, (\ref{eq:gamma}) implies for $0<\rho <w_0$,
$$\zeta_K(1)={-1 \over 2\pi i}\oint\limits_{|z|=\rho}{dz \over \hat{t}(z)z^K}=-Res\left({1 \over \hat{t}(z)z^K}; w_0 \right)-{1 \over 2\pi i}\oint\limits_{|z|=\gamma}{dz \over \hat{t}(z)z^K}.$$
Observe that $Res\left({1 \over \hat{t}(z)z^K}; w_0 \right)$ is a constant multiple of ${1 \over w_0^K}$ since $w_0$ is a root of $\hat t(z)$ of algebraic multiplicity one. 
Thus, since $w_0 <\gamma$ and
$~\left|{1 \over 2\pi i}\oint\limits_{|z|=\gamma}{dz \over \hat{t}(z)z^K}\right| \leq {1 \over \gamma^K \min\limits_{|z|=\gamma}|\hat{t}(z)|}$,

$${\zeta_{K+1}(1) \over \zeta_K(1)}=\left|{\zeta_{K+1}(1) \over \zeta_K(1)}\right| \rightarrow {1 \over w_0} \quad \text{ as } K \rightarrow \infty.$$
Hence Proposition \ref{main} will imply the following lemma.
\begin{lem} \label{maincor1}
Suppose $~\limsup\limits_{j \rightarrow \infty} T_j^{1/j} <\infty$. 
Then, for each positive integer $j$, the limit 
$$\xi(j)=\lim\limits_{K \rightarrow \infty}\xi_K(j)$$ exists, and $\xi(j)=w_0^{j-1}$.
\end{lem}

\medskip
\noindent
The converse is also true. Specifically, suppose the limit
$$R=\lim\limits_{k \rightarrow \infty} {\zeta_{k+1}(1) \over \zeta_k(1)}$$
exists and is finite. Then, since $~\zeta_k(1) \geq T_j^{k/j}~$ for all $j \in \mathbb{N}$ and $k \in j\mathbb{N}$,
$$\limsup\limits_{j \rightarrow \infty} T_j^{1/j} \leq \lim\limits_{k \rightarrow \infty} \big[\zeta_k(1)\big]^{1/k} = R < \infty.$$
Hence, we have proven another lemma. 

\begin{lem} \label{maincor2}
Suppose $~\limsup\limits_{j \rightarrow \infty} T_j^{1/j} =\infty$. 
Then, the limit 
$~\lim\limits_{K \rightarrow \infty}\xi_K(j)~$ 
does not exist at least for some $j$.
\end{lem}

We now combine the results in Lemmas~\ref{maincor1} and \ref{maincor2} into
the following theorem.
\begin{thm}
\label{thm_cor}
Suppose $~\limsup\limits_{j \rightarrow \infty} T_j^{1/j} <\infty$. 
Then, for each positive integer $j$
$$\xi(j)=\lim\limits_{K \rightarrow \infty}\xi_K(j)
=\lim\limits_{K\to\infty}\frac{\cN_{j}[K]}{\cN_1[K]}=R^{1-j},$$ 
where $1/R=w_0$ is the only real root of the function
$\hat{t}(z)=-1+2z+\sum\limits_{j=1}^\infty z^j T_j$ 
in the interval $\left(0,{1 \over 2}\right]$. 
Conversely, if 
$~\limsup\limits_{j \rightarrow \infty} T_j^{1/j} =\infty$, 
then the limit 
$\lim\limits_{K \rightarrow \infty}\xi_K(j)=\lim\limits_{K\to\infty}\frac{\cN_{j}[K]}{\cN_1[K]}$ 
does not exist at least for some $j$.
\end{thm}

We notice that the fact that $R$ is reciprocal to the solution
of $\hat t(z)=0$ was noticed by Peckham \cite{Pec95}, under the
assumption $N_k \sim c\,R^{K-k}$, as $K\to\infty$.
Below we give several examples of Theorem~\ref{thm_cor}.

\bigskip
\noindent
{\bf Example 1: Shallow side-branching.}
Suppose $T_k=0$ for $k\ge 3$, that is we only have 
``shallow'' side-branches of orders $\{j-2,j\}$ and $\{j-1,j\}$.
Then
\[\hat t(z) = -1 + (T_1+2)\,z + T_2\,z^2.\]
The only root of this equation within $[0,1/2]$ is
\[w_0 = \frac{\sqrt{(T_1+2)^2+4T_2}-(T_1+2)}{2\,T_2},\]
which leads to
\[R=\frac{1}{w_0}=\frac{\sqrt{(T_1+2)^2+4T_2}+(T_1+2)}{2}.\]
In particular, if $T_k=0$ for $k\ge 2$, then $R=T_1+2$;
such trees are called ``cyclic'' \cite{Pec95}.
This shows that the entire range of Horton exponents $2\le R < \infty$
can be achieved by trees with only very shallow side-branching.
This also shows that $T_1\ge1$ leads to $R\ge 3$,
which seems to be the case for most of the observed branching systems.

\bigskip
\noindent
{\bf Example 2: Tokunaga self-similarity.} 
Suppose $T_j=a\,c^{j-1}$, where $a,c>0$, as in \cite{Tok78, Pec95,MG08}. Then
$$\hat{t}(z)=-1+2z+az \sum\limits_{j=1}^\infty (cz)^{j-1}=-1+2z+{az \over 1-cz}={-1+(a+c+2)z-2cz^2 \over 1-cz}.$$
Here
$$-{1 \over \hat{t}(z)}={1-cz \over 1-(a+c+2)z+2cz^2},$$
and the discriminant is positive,  $(a+c+2)^2-8c > (c+2)^2-8c=(c-2)^2 \geq 0$. Therefore, there will be two positive roots, $p_1>p_2$ of the denominator $1-(a+c+2)z+2cz^2$, and
$$-{1 \over \hat{t}(z)}={1-cz \over 2c(p_1-p_2)(z-p_1)}-{1-cz \over 2c(p_1-p_2)(z-p_2)}.$$
Thus, since ${1 \over z-p}=-\sum\limits_{k=0}^\infty {1 \over p^{k+1}}z^k$ for $|z|<|p|$, formula (\ref{zetaT}) implies
\begin{equation} \label{HortonZeta}
\zeta_{K+1}(1)={1 \over 2c(p_1-p_2)}\left({1-cp_2 \over p_2^{K+1}} -{1-cp_1 \over p_1^{K+1}} \right)
\end{equation}
for $|z|$ small enough, where one can easily check that $1>cp_2$. Therefore the conditions of Proposition \ref{main} are satisfied with 
$${1 \over R}=p_2={a+c+2 -\sqrt{(a+c+2)^2-8c} \over 4c}.$$
Hence,
\begin{equation} \label{HortonR}
R={a+c+2 +\sqrt{(a+c+2)^2-8c} \over 2}
\end{equation}
as in \cite{Tok78,Pec95,MG08}. Also, in agreement with the Lemma \ref{maincor1}, $w_0=p_2={1 \over R}$.

\bigskip
\noindent
{\bf Example 3: ``Differentiated Tokunaga" self-similarity.} 
Suppose $T_j=a \cdot jc^{j-1}$, where $a,c>0$. Then
$$\hat{t}(z)=-1+2z+az \sum\limits_{j=1}^\infty j (cz)^{j-1}=-1+2z+{az \over (1-cz)^2}={2c^2z^3-c(c+4)z^2+(a+2c+2)z-1 \over (1-cz)^2}.$$
Here ${1 \over R}=w_0$ is the smallest positive real root of polynomial
$$p(z)=2c^2z^3-c(c+4)z^2+(a+2c+2)z-1.$$
Now, since $\hat{t}(1/2)>0$, $~w_0 \in \left(0,{1 \over 2}\right)$. 
In this example, we cannot derive an explicit formula for $R=R(a,c)$.
However, we solve $p(w_0)=0$ for $c>0$, obtaining the following relation
among $a,c$ and $R$:
$$c={1 \over w_0}+\sqrt{a \over (1-2w_0)w_0}=\left(1+\sqrt{a \over R-2}\right)R.$$

%

\bibliographystyle{amsplain}

\end{document}